\documentclass[10pt,letterpaper]{article}
\usepackage{opex3}
\usepackage{color}
\usepackage{cite}
\usepackage{epsfig}
\usepackage{amssymb,amsmath,amsthm,graphics}
\usepackage{epstopdf}
\usepackage[colorlinks=true,urlcolor=blue,linkcolor=black,citecolor=black,breaklinks=true,bookmarks=false]{hyperref}

\begin{document}

\title{Optical textures: characterizing spatiotemporal chaos}

\author{Marcel G. Clerc,$^1$  Gregorio Gonz\'alez-Cort\'es,$^1$ Vincent Odent,$^{1,2}$ and Mario Wilson $^{1,3,*}$}

\address{$^1$Departamento de F\'isica, Facultad de Ciencias F\'isicas y 
Matem\'aticas, Universidad de Chile, Blanco Encalada 2008, Santiago, Chile\\
$^2$Present address: Universit\'e Lille 1, Laboratoire de Physique des Lasers, Atomes et
Mol\'ecules, CNRS UMR8523, 59655 Villeneuve d'Ascq Cedex, France\\
$^3$CONACYT -- CICESE, Carretera Ensenada-Tijuana 3918,
Zona Playitas, C.P. 22860, Ensenada, M\'exico}

\email{$^*$mwilson@cicese.mx} 

\begin{abstract}
Macroscopic systems subjected to injection and dissipation of energy can
exhibit complex spatiotemporal behaviors as result of dissipative self-organization. 
Here, we report a one and 
two dimensional  pattern forming set up, which 
exhibits a transition from stationary patterns to spatiotemporal
chaotic textures,  based on a nematic liquid crystal layer with 
spatially modulated input beam and optical feedback. Using an adequate projection of spatiotemporal diagrams, 
we determine the largest Lyapunov exponent. Jointly, 
this exponent and Fourier transform  allow us to distinguish between spatiotemporal chaos
and amplitude turbulence concepts, which are usually merged.
\end{abstract}



\section{Introduction}
\begin{figure*}
\centering\includegraphics[width=12cm]{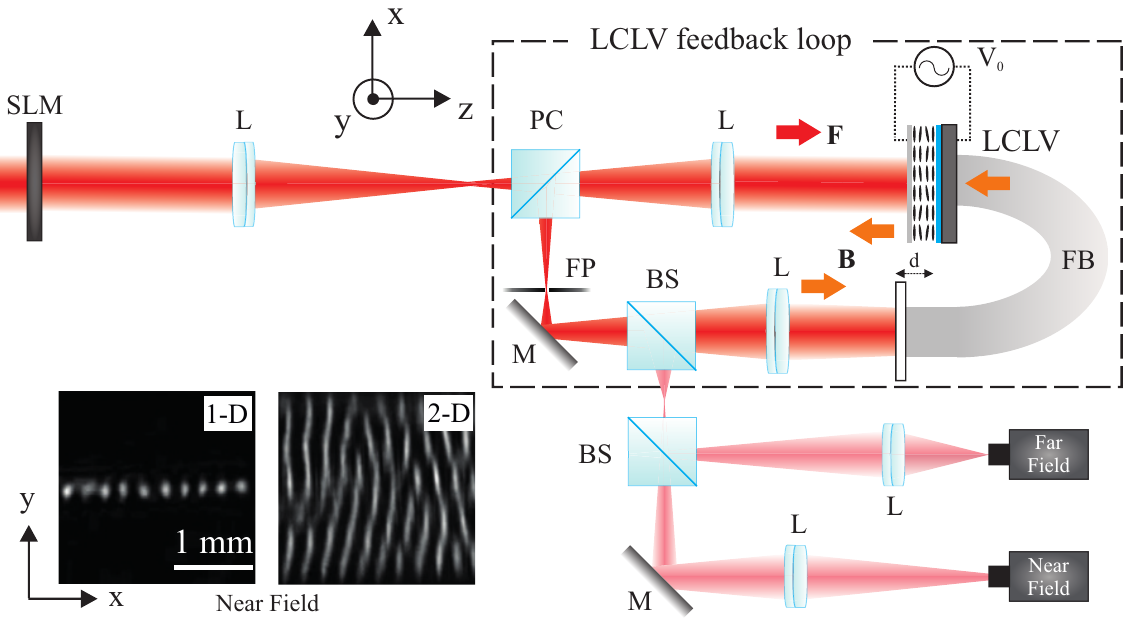} 
\caption{Schematic representation of the experimental setup. LCLV stands for the liquid crystal light valve, L represents 
the achromatic lenses with a focal distance $f=25\:cm$, M are the mirrors, 
FB is an optical fiber bundle, BS stands for the beam splitters, PC 
represents a polarizing cube, SLM is a spatial light modulator and FP represents the Fourier plane.
F and B stand for the forward (incoming) and backward (reflected) beam respectively. $d$ is the equivalent optical length. 
In the bottom left  part of the image, two examples of obtained patterns  and textures.}
\label{Fig1}       
\end{figure*}

Optical  systems maintained far from equilibrium,
through the injection and dissipation  of energy, can
present  spatiotemporal  structures, 
{\it patterns}~\cite{Arecchi90,Huyet,Tredicce96,Mamaev98,Rogers2004,ResidoriReport,PatternsOptics,Descalzi2011,mustapha}.
These structures appear as a way to 
optimize energy transport and momenta~\cite{Prigogine}.  
Patterns are the result of the interplay between the linear gain and the
nonlinear saturation mechanisms. In many physical systems,
these structures are stationary and emerge as a spatial instability of a uniform state
when a control parameter is changed and surpasses a critical
value, which usually corresponds to imbalances of forces.
As the parameters of the system are changed, stationary patterns can become 
unstable and bifurcate to more
complex patterns, even into aperiodic dynamics states 
\cite{Nicolis,DefectMediated,DefecMediated1,GorenEckmannProcaccia}.  
This behavior is characterized by complex spatiotemporal dynamics exhibited by the 
pattern and a continuous coupling between spatial modes in time. 
Complex spatiotemporal dynamics  of patterns have been observed, for example, 
in fluids \cite{Bodenschatz,Bodenschatz1,Bodenschatz2,Miranda,Limat}, chemical reaction-diffusion systems \cite{Flesselles}, 
cardiac fibrillation \cite{Weiss}, electroconvection \cite{Zhou}, fluidized granular matter \cite{Moon},  
nonlinear optical cavities \cite{Arecchi90,Huyet,Tredicce96,Mamaev98} 
and in a liquid crystal light valve  \cite{ResidoriReport, VBCR2013,VBCR20131}.
In most of these studies, complex behaviors are characterized by spatial and temporal 
Fourier transforms, wave vector distribution, 
filtering spatiotemporal diagrams, power spectrum of spatial mode, 
length distributions, Poincar\'e maps and number of defects as a function of the parameters.
However, in these experimental studies, spatiotemporal complexity has 
not been characterized using rigorous tools of 
dynamical systems theory as Lyapunov exponents \cite{Manneville90,pastur}. 
These exponents characterize the exponential sensitivity 
of the dynamical behaviors under study and in turn gives a characteristic time scale on which one has 
the ability to predict the time evolution of the system. When the largest Lyapunov exponent (LLE)
is positive (negative) the system under study is chaotic (stationary).
Recently, we have computed the experimental LLE and characterized the
spatiotemporal chaos in two spatial dimensions in a  liquid crystal light valve (LCLV) 
with optical feedback~\cite{CGW2015}. The dynamics exhibited by the LCLV with 
optical feedback is characterized by the changes that 
molecular orientation induces in the phase of the reflected light,  
which, in its turn---optical feedback---produces a voltage that 
reorients the liquid crystal molecules.  

One of the most important concepts in complex spatiotemporal dynamics is turbulence: the power spectrum as a result of the transport of some physical quantity of different scales shows a power law decay as its main signature \cite{Frisch}.
The aim of this article is to investigate one and two dimensional pattern forming system in a LCLV, shown in Fig.~\ref{Fig1},
which exhibits a transition from stationary patterns to spatiotemporal chaotic textures and to quasiperiodicity. 
Using  adequate spatiotemporal diagrams, we obtain the LLE. Jointly, this exponent and Fourier transform allow us to distinguish between spatiotemporal chaos and amplitude turbulence concepts, which are usually merged.

\section{Experimental details}
The liquid crystal light valve with an optical feedback is a flexible optical experimental setup 
that exhibits pattern formation \cite{ResidoriReport}  [see Fig.~\ref{Fig1}]. 
The LCLV is illuminated by an expanded and collimated He-Ne laser beam, $\lambda=633 \:nm$, 
 with $3$~cm transverse diameter and power $I_{in}=6.5$ mW/cm$^2$, linearly polarized along the vertical axis. 
 Once shone into the LCLV, the beam is reflected by the dielectric mirror 
deposed on the rear part of the cell and, thus, sent to the polarizing cube. Due to the phase-change the light suffers in the reflection, 
the polarizing cube will send the  reflected light into the feedback loop. To close the feedback 
 loop, a mirror and an optical fiber bundle  are used, these elements 
 assure the light to reach the photoconductor placed in the back part of the LCLV. 
In the feedback loop, a 4-f array is placed in order to obtain a self-imaging 
configuration and access to the Fourier plane, this array is constructed 
with 2 identical lenses  with focal length $f=25~cm$ placed in such a way 
that both sides of the LCLV are conjugated planes. We filter the Fourier plane in order to force the system to exhibit roll-patterns in a given direction.
Thanks to this configuration the free propagation length in the feedback loop 
can be easily adjusted. For the performed experiments an optical equivalent length of $d=-4$~cm was used.
A spatial light modulator (SLM) was placed in the input beam optical path with a $1:1$ imaging between 
the SLM and the frontal part of the LCLV.    
With the aid of a specialized software, a square mask was produced and sent to the 
SLM. The SLM and the polarizing cube combination allow to impose an arbitrary shape 
to the input beam.  
For a uniform mask  of 160 gray-value, the typical input intensity would be 
$I_{w}=0.83\:mW/cm^2$. To obtain the shape used in the experiments, 
one and two-dimensional masks, $I(x,y)$, were created and, by means of these masks, one and two dimensional patterns 
can be obtained as can be seen in the bottom left part of Fig.~\ref{Fig1}. 
The system dynamics is controlled by adjusting the external voltage $V_0$ 
applied to the LCLV.

\section{From stationary to disordered dynamics}
\begin{figure*}
\centering\includegraphics[width=7cm]{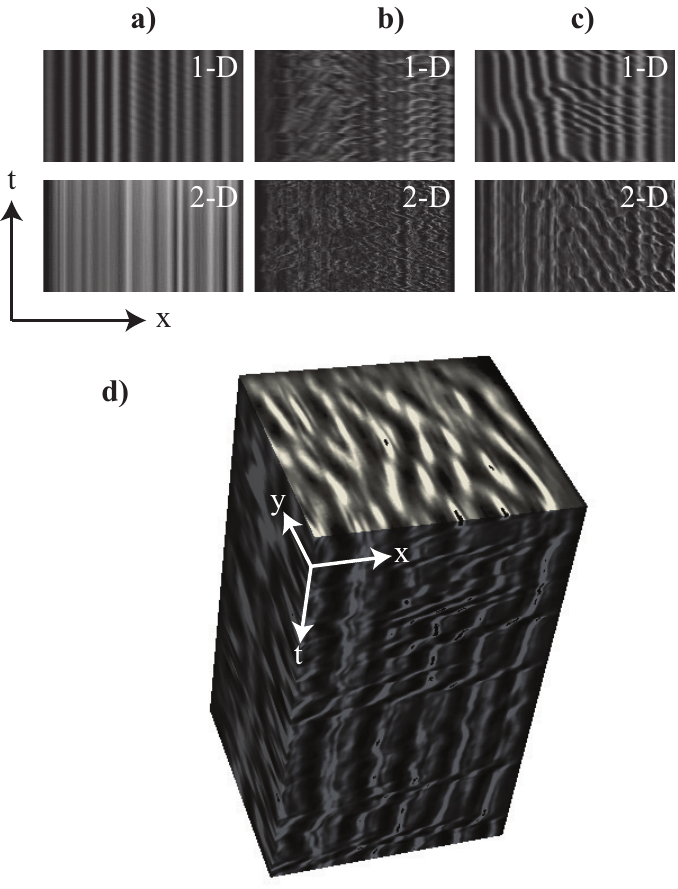} 
\caption{Spatial textures in LCLV with optical feedback at different voltage $V_0$. Top panels 
correspond to  spatiotemporal diagrams of observed dynamics working with one-dimensional patterns. 
Middle panels panels stand for projected spatiotemporal diagrams of two-dimensional textures. 
 a) Periodic regime, b) chaotic behavior, c) quasi-periodic dynamics, 
and d) a 3-D spatiotemporal diagram of a complex texture found in the intermittent regime at $V_0=4.3V{rms}$. All values were taken during $100~s$ and are normalized.}
\label{Fig2}       
\end{figure*}

The presented dynamics in the LCLV 
have been explored in two different configurations, the first one using an intensity 
mask of zero-level intensity everywhere except for a central square part with length $a_0=2.5$ mm (2-D 
mask), and the second one with a zero-level intensity except on a narrow channel of 150 $\mu$m width and $2.5$ mm
length (1-D mask). 
The injected intensity  is spatially modulated as $I_{in}=I_0(x,y)$, where $I_0$  can be controlled by changing the mask created 
in the SLM, and
 $\{x, y\}$ are the transverse coordinates of the sample. $I_0$ 
is measured when imposing a given gray-value to the illuminated area, that is, for the 2-D mask
\[
I_{0}(x,y)=\left\{
\begin{array}
[c]{c}
I_{0}+b_{0} \ \  \ \ \ \  |x| \leq a_0, \ {\text and }  \ \  |y| \leq a_0\\
b_{0}  \ \ \ \  \ \ \ \ \ \  \ \ \ \ \ \  \ \ \ \ \ \  \ \ \ \ else
\end{array}
\right.
\]
when $b_0$ is constant throughout the sample and $|x|>0$, $|y|>0$. The same applies to 1-D mask 
with the only difference that $ |y| = 150\: \mu$m, which is small enough, compared with the 
pattern wavelength, to neglect its size and consider it as a 1-D mask. In the presented configurations 
$I_0 =0.9\: mW/cm^2$ and $b_0=0.1\:mW/cm^2$.  The alternating voltage  $V_0$ has been varied between $3$ and $7$ $V_{rms}$, 
at a constant frequency $f_0=5\:kHz$,
 starting with the appearance of stationary roll-patterns.
For different $V_0$ values, the dynamical behavior obtained 
in the system was recorded with a CCD  camera.
Figure~\ref{Fig2} shows the spatiotemporal 
evolution of the observed patterns in one and two dimensions, respectively. 
This evolution is characterized by projected 
spatiotemporal diagrams,  which are constructed, in the 2-D experiments, 
by picking an arbitrary line---transversal to the 
rolls 
direction---in the illuminated zone and superposing it as time evolves; 
in the 1-D experiments this  construction is simpler, is enough to superpose the pattern as the time evolves. 
The system exhibits stationary  stripe patterns [cf. Fig.~\ref{Fig2}(a)].
These patterns are  induced by a spatial filtering in the Fourier plane [cf. FP in Fig.~\ref{Fig1}].  
Actually, through a slit, we can filter spatial modes.

Increasing $V_0$, the dynamics shown by the pattern becomes 
abruptly complex [cf. Figs.~\ref{Fig2}(b) and \ref{Fig2}(d)]. Clearly, in the projected spatiotemporal diagram, we detected an intermittent behavior. 
That is, the pattern exhibits aperiodic oscillations invaded by large fluctuations, generating several 
spatial and temporal dislocations. Likewise, the system exhibits a high spatiotemporal complexity. 
This kind of disorder is usually associated to spatiotemporal chaotic textures
\cite{CGW2015,Nicolis1995,Verschueren2013,Bodenschatz2002}.
Figure \ref{Fig2}(d) shows a 3-D spatiotemporal diagram, 
from this diagram it is clear that an arbitrarily chosen line represents the dynamics.

Further increasing $V_0$,  the pattern begins to oscillate in a complex manner [see Fig.~\ref{Fig2}(c)]. 
We observe in the projected spatiotemporal diagram  local
waves, oscillations and spatiotemporal dislocations.
Similar dynamics has been reported in one-dimensional inhomogeneous 
systems \cite{ClercOrtega2014,Louvergneaux2001,Szwaj2005}.
In our experiments these inhomogeneities can be caused by the 
inherent imperfections and inhomogeneities 
induced by the filter in the Fourier plane. Hence, this kind of dynamical 
behavior could be expected.
The complex dynamics exhibited by this pattern 
is constantly repeated over time. Which leads us to infer that this kind of behavior 
could be quasiperiodicity.

\begin{figure*}
\centering\includegraphics[width=7cm]{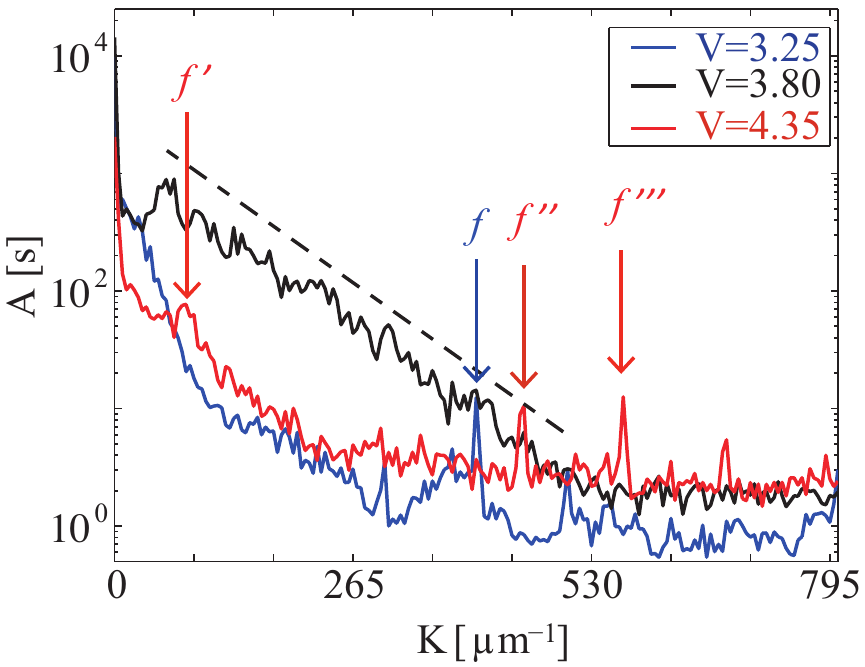} 
\caption{Fourier spectra for three different dynamical regimes. In gray (blue) the Fourier spectrum 
of a stationary pattern, in light-gray (red) a quasi-periodic regime, it can be observed the emergence of 
incommensurable frequencies ($f'$, $f''$ and $f'''$). In black, a broadened spectrum that corresponds 
to a chaotic texture, presenting an exponential decay of the modes marked by the dashed line.}
\label{Fig3}       
\end{figure*}

A mathematical tool for analyze the spatial modes interaction is the Fourier spectrum.
Figure \ref{Fig3} shows the Fourier spectra of  different dynamical regimes. Showing that the 
dynamics changes between stripe patterns, quasi-periodicity and spatiotemporal chaotic textures. 
The stationary pattern is characterized 
by a dominant wavelength $f$. The width of this peak is due to temperature fluctuations and 
dynamics of defects such as dislocations and boundary grains. The quasi periodic texture is characterized by the 
appearance of   incommensurable wavelengths, $\{f',f'''\}$, with respect to the main wavelength $f''$ and its harmonics. 
The spatiotemporal  chaotic texture is characterized by presenting an enlarged spectrum 
as a result of the interaction between the  main  incommensurable modes \cite{Kuramoto84}. 
Note that in this regime, the modes are coupled with 
exponential decay [see the dashed line in Fig.~\ref{Fig3}]. Therefore, 
the system does not exhibit power spectrum behavior which is the hallmark 
of turbulence dynamics \cite{Frisch}.

\section{Quantifying the dynamics}
A characterization of complex dynamics 
like chaos and spatiotemporal chaos
can be done by means of Lyapunov exponents. There are as many exponents
as the dimension of the system under study. 
The analytical study of Lyapunov exponents is a paramount 
endeavor and in practice inaccessible, then the pragmatic strategy 
is a numerical derivation of the exponents. 
From  experimental data, in the case of low-dimensional dynamical systems, 
by means of recognition 
of initial conditions one can determine the LLE \cite{Swinney85}.
This exponent accounts for the greatest exponential growth
and its defined by
\begin{equation}
\lambda_0= \lim_{t\rightarrow \infty} \lim_{\Delta_0 \rightarrow 0} \frac{1}{t} 
\ln\left[\frac{\left|\left| {{\bf u}(x,t)-{\bf u}'(x,t)}\right|\right| }{\left|\left| {{\bf u}(x,t_o)-{\bf u}'(x,t_o)}\right|\right|}\right],
\end{equation}
where ${\bf u}(x,t)$ and ${\bf u}'(x,t)$ are given fields, $\Delta_o \equiv||{\bf u}(x,t_0)-{\bf u}'(x,t_0)||$
 and $ ||f(x,t) ||^2 \equiv \int |f(x,t)|^2 dx$
is a norm. $\Delta(t) \equiv||{\bf u}(x,t)-{\bf u}'(x,t)||$ stands for the global evolution of the difference between the fields.

\begin{figure*}
\centering\includegraphics[width=8cm]{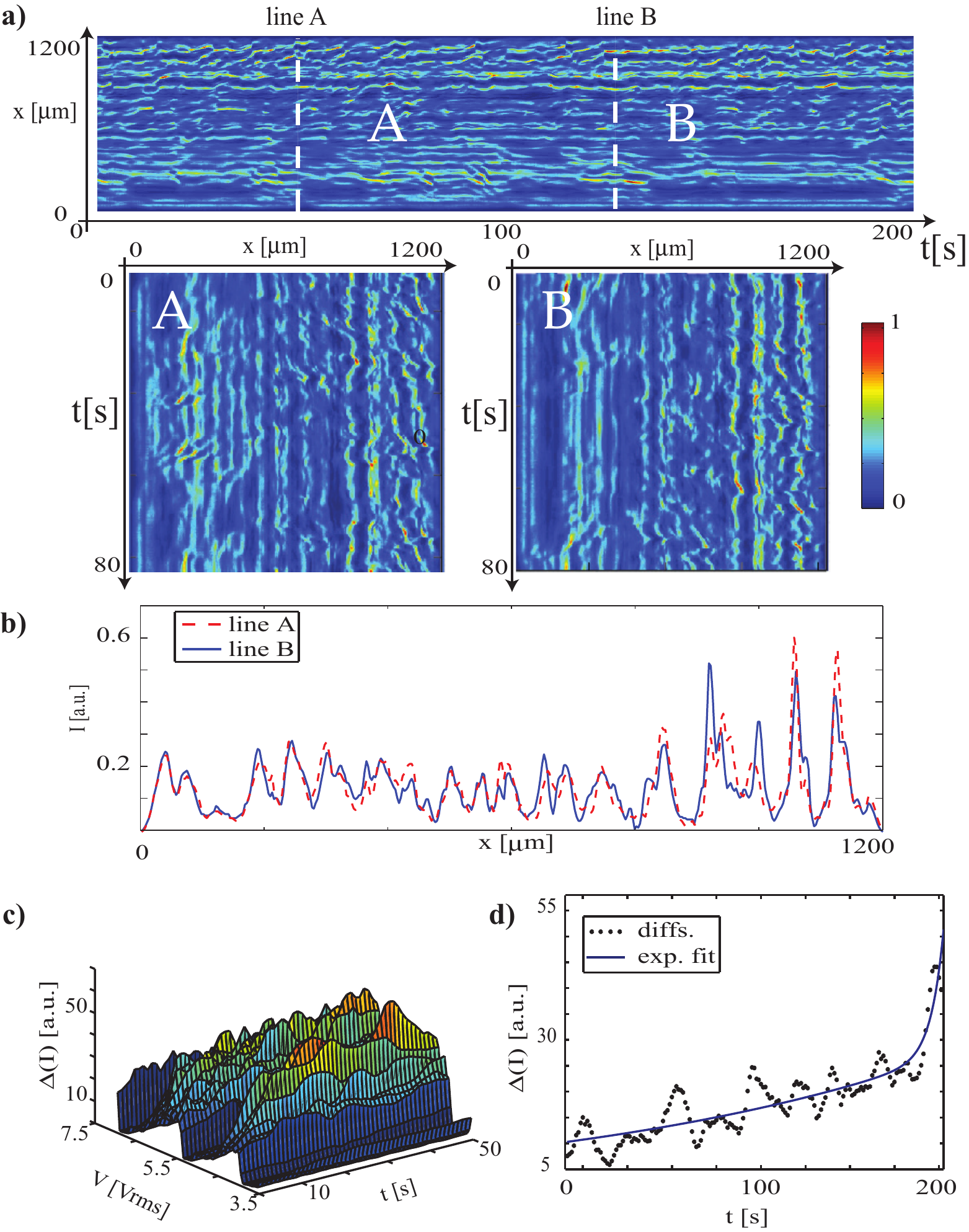} 
\caption{LLE estimation for the LCLV with optical feedback with 
$V_0=4.70V_{rms}$. a) Projected spatiotemporal diagram of the LCLV. 
The bottom panels account for spatiotemporal diagram with  close initial conditions (lines $1$ and   $2$), b) 
Intensity profiles of lines 1 and  2, c) A 3-D graph  that shows evolution of differences $\Delta(t,V_0)$ for different applied voltages 
$V_0$, it can be noted that not always the trajectories are exponentially separated, and d) Temporal evolution of global 
difference $\Delta(t)$, dots stand for experimental data and the continuous curve is the exponential fitting $\Delta(t) 
\approx ae^{bt}+ce^{dt}$ with $a=0.0045$, $b=8.813$, $c=16.57$ and $d=0.2773$.}
\label{Fig4}       
\end{figure*}

When $\lambda_0$ is positive or negative, the perturbation of a given trajectory 
is characterized by an exponential  separation or approach, respectively.
Dynamical behaviors with zero LLEs correspond to equilibrium with invariant directions, 
such as periodic or quasi-periodic solutions and non-chaotic attractors 
\cite{Ott2002}. Hence, the LLE is an exceptional order parameter for characterizing  
transitions from stationary to complex spatiotemporal dynamics.

 Experimentally, to estimate the LLE,  it is  mandatory to have two close initial conditions
and observe if their evolution diverge at large times \cite{CGW2015}. 
The implemented method needs, as a first step,  to find two close fields 
[see lines 1 and 2 in Figs.~\ref{Fig4}(a) and  \ref{Fig4}(b)] 
along the projected  spatiotemporal 
diagrams and compute their difference $\Delta_0$. The temporal evolution of the difference should be 
given by $\Delta(t) \approx  \Delta_0 e^{\lambda_0 t}$ for large $t$ [cf  Figs.~\ref{Fig4}(c) and \ref{Fig4}(d)]. 
Due to the complexity of evolution of the difference between fields---clearly the number of positive Lyapunov 
exponents is huge---we will consider at least two unstable growth directions,
that is  $\Delta(t) \approx ae^{bt}+ce^{dt}$ [cf. Fig.~\ref{Fig4}].

\begin{figure*}
\centering\includegraphics[width=8cm]{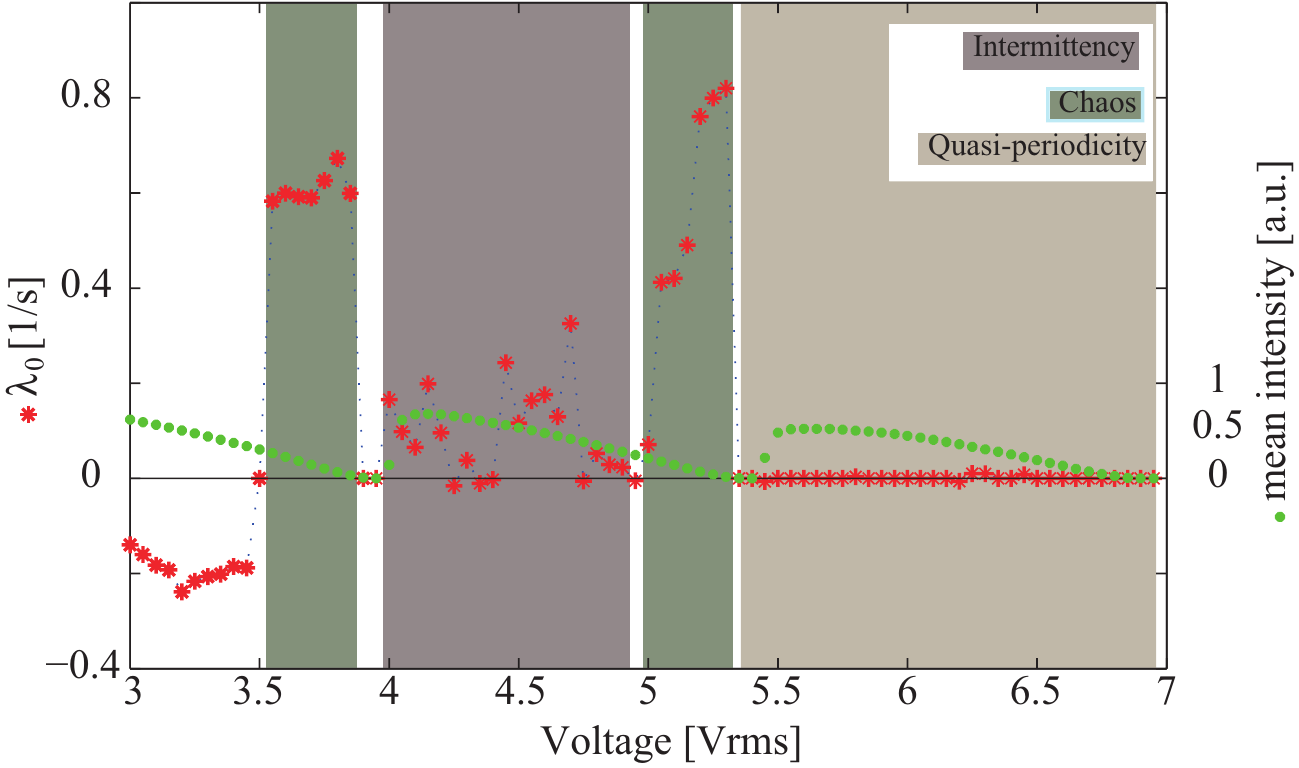} 
\caption{Bifurcation diagram constructed with the estimated LLE as a function of applied 
voltage $V_0$ of the LCLV with optical feedback. The diagram is clearly separated in four dynamical 
regimes, stationary patters before $V_0=3.5V_{rms}$, chaos shadowed in gray (green), intermittency between chaos and quasi-periodicity in dark gray (gray) area and a window of quasi-periodicity shadowed at the end in light gray. The stars correspond to the calculated 
LLE while the circles show the normalized mean intensity present in the LCLV. }
\label{Fig5}       
\end{figure*}

A bifurcation diagram was constructed with the obtained LLEs as can be seen in Fig.~\ref{Fig5}. 
The system starts with stationary stripe patterns at $V_0=3.0\:V_{rms}$ and the dynamics remains unchanged until the 
applied voltage reached $V_0=3.5\:V_{rms}$. At  this voltage, the LLE goes to zero, meaning that the system exhibits a bifurcation.
Experimentally, we observed that the steady pattern changes to an aperiodic 
regime. The chaotic behavior remains until 
the mean intensity in the LCLV 
destroys  the chaotic attractor due to destructive interference at $V_0=3.9\:V_{rms}$ [see 
Fig.~\ref{Fig5}], 
causing a crisis. Once the light is recovered, the system enters in an intermittent regime between 
chaos and quasi-periodicity. After this window the system 
becomes chaotic until the attractor is annihilated by destructive interference  
at $V_0=5.35 \:V_{rms}$. Once the light is recovered the 
system remains in a quasi-periodic regime until the next cycle of destructive interference arrives.   
This dynamic regime is characterized by having an oscillatory pattern which LLE is zero.

Given a temporal signal, the attractor of the system can be built by taking the signal 
at different periodic times (arbitrary periodic separation $\tau$)  and constructing the vector 
$\{I(x,t),I(x,t+\tau,x),I(x,t+2\tau),\cdots\}$
with $x$ as a fixed position, {\it phase space 
reconstruction} \cite{Abarbanel}. 
The three different attractors that can be reconstructed using this embedding method are a fixed point, a torus and a strange attractor.
For low voltage a fixed point can be seen (stationary pattern).
Increasing the tension $V_0$, the phase space 
reconstruction exhibits a
torus (quasi-periodic pattern) and  strange attractor (spatiotemporal texture)\cite{CGW2015}.

\section{Conclusions}
Our study provides clear evidence that the LCLV with optical feedback is
spatiotemporally chaotic in a certain range of parameters. The LLEs are experimentally accessible and allow us
to characterize  the transitions from stationary to complex spatiotemporal dynamics.
Certainly new concepts in the theory of dynamical systems must be 
developed to achieve a better experimental characterization of spatiotemporal 
complex behaviors. Notwithstanding, jointly the LLEs and power spectrum allow us
distinguishing well-established dynamical behaviors such as amplitude turbulence and spatiotemporal chaos, 
which are often merged and confused.

\section*{Acknowledgments} MGC and MW acknowledge the support of FONDECYT N$^{\circ} 1150507$ and N$^{\circ} 3140387$ respectively. 
VO acknowledges the support of the {\lq}R\'egion Nord-Pas-de-Calais'.

\end{document}